# In vivo measurement of human brain elasticity using a light aspiration device

Schiavone P.[1-2], Chassat F.[3], Boudou T.[1], Promayon E.[1], Valdivia F.[4], Payan Y.[1]

1 Laboratoire Techniques de l'Ingénierie Médicale et de la Complexité - Informatique, Mathématiques et Applications de Grenoble, UMR CNRS 5525 - Université Joseph Fourier, Pavillon Taillefer, Faculté de Médecine, Domaine de la Merci, 38706, La Tronche cedex, France

2 Laboratoire des Technologies de la Microélectronique CNRS, c/o CEA Grenoble, 17 rue des Martyrs, 38054 Grenoble cedex 09, France

3 Centro de Modelamiento Matemático, Av. Blanco Encalada, 2120 piso 7, Santiago, Chile

4 Instituto de NeuroCirugía Asenjo, PO Box 3517, Santiago, Chile

# Abstract

The brain deformation that occurs during neurosurgery is a serious issue impacting the patient "safety" as well as the invasiveness of the brain surgery. Model-driven compensation is a realistic and efficient solution to solve this problem. However, a vital issue is the lack of reliable and easily obtainable patient-specific mechanical characteristics of the brain which, according to clinicians' experience, can vary considerably. We designed an aspiration device that is able to meet the very rigorous sterilization and handling process imposed during surgery, and especially neurosurgery. The device, which has no electronic component, is simple, light and can be considered as an ancillary instrument. The deformation of the aspirated tissue is imaged via a mirror using an external camera. This paper describes the experimental setup as well as its use during a specific neurosurgery. The experimental data was used to calibrate a continuous model. We show that we were able to extract an *in vivo* constitutive law of the brain elasticity: thus for the first time, measurements are carried out per-operatively on the patient, just before the resection of the brain parenchyma. This paper discloses the results of a difficult experiment and provide for the first time in-vivo data on human brain elasticity. The results point out the softness as well as the highly non-linear behavior of the brain tissue.

# 1 Introduction

Computer assisted and image-guided neurosurgery is constantly evolving, from the frame-based stereotaxy introduced more than twenty years ago [Kelly, 1986] to the frameless neuronavigation systems, developed in the past ten years, and now routinely used in standard neurosurgical procedures. During neuronavigation, pre-operatively acquired image data of different modalities such as Computed Tomography (CT) or Magnetic Resonance Imaging (MRI) are registered to the patient's anatomy in the physical space of the operating room. As a consequence, the physical locations of surgical ancillaries are mapped onto their corresponding positions within the image data. Whereas rigid registration is relatively straightforward, clinical studies in image guided neurosurgery have highlighted limitations of this approach during the surgery. Indeed, especially in the case of large skull openings, deformations of the brain tissue may occur due to known and unknown physical phenomena (e.g. dura opening, intra-operative position of the patient, loss of cerebrospinal fluid, surgical manipulation, characteristics of the tissue, etc) and to known and unknown physiological phenomena (e.g. swelling due to osmotic drugs, anaesthetics). Studies have reported a continuous dynamic shift of the brain tissue evolving differently in distinct brain regions, with a surface shift (up to 24 mm) that occurs throughout surgery and with a subsurface shift (exceeding 3 mm for the deep tumour margin) that mainly occurs during resection [Hastreiter et al. , 2004] [Nabavi et al. , 2001]. As a consequence of this phenomenon, known as "brain-shift", the pre-operatively acquired images are no longer a valid representation of the patient's status. Therefore, the accuracy of any neuronavigation system significantly deteriorates during the ongoing course of the surgery.

To address this problem, researchers and clinicians introduced computational models in order to compensate brain shift by updating the pre-operative images according to measured intra-operative brain deformation. The first algorithms attempted to deform the pre-operative images using image-based models. Different non-rigid registration methods were therefore provided to match intra-operative images (mainly MRI exams) with pre-operative ones [Hata et al. , 2000][Hastreiter et al. , 2004][Shattuck & Leahy, 2002]. More recently, biomechanical models of the brain tissue were used to constrain the image registration: they can infer a volumetric deformation field in the images to register from matched contours [Kyriacou et al. , 1999][Hagemann et al. , 1999] and/or surfaces [Ferrant et al. , 2002] [Clatz et al. , 2005]. Arguing against the high cost of the intra-operative MRI imaging devices, some authors have proposed to couple the biomechanical model of the brain with low-cost readily available intra-operative data such as laser-range scanner systems [Audette et al. , 2003][Miga et al. , 2003] or intra-operative ultrasound [Comeau et al. , 2000][Bucki *et al.* , 2007][Reinertsen *et al.* , 2007]. These techniques give a crucial and very central position to the patient biomechanical brain model. This implies that a lot of effort is needed to design the brain model as well as to validate the simulation results regarding to clinical data. Another key aspect is the estimation of the mechanical behaviour of the brain soft tissue. For most models that assume an elastic formulation to describe brain tissue, the estimation of the constitutive law of the material, i.e. the relationship between strain and stress, has to be established. In this framework, the rheological experiments of Miller significantly contributed to the understanding of the mechanics of the brain tissue [Miller, 2002]. His extensive investigation in brain tissue engineering showed very good concordance of a hyper-viscoelastic constitutive equation with in vivo (for swine under anaesthesia [Miller et al. , 2000]) and in vitro [Miller & Chinzei, 2002] experiments. Because of the challenges presented by model-updated image-guided neurosurgery in terms of computational time associated with the biomechanical models, most authors propose to simplify the modelling hypothesis by relying on a small deformation linear elastic constitutive equation to characterize the mechanical behaviour of the brain tissue. Even in this quite restrictive framework, parameter values used to define the mechanical behaviour vary significantly, with a Young modulus between 0.6 kPa [Clatz et al., 2005] and 180 kPa [Kyriacou et al., 1999], and Poisson's ratio between 0.4 [Skrinjar et al., 2002] and 0.499 [Miller et al., 2000].

This paper aims at demonstrating the feasibility of a simple device, able to provide patient specific estimations of the brain elasticity that can meet the very severe sterilization and handling issues imposed during surgery. We show that we were able to extract a constitutive law of the brain elasticity obtained *in vivo*, through intra-operative measurements that are, for the first time, carried out per-operatively just before the resection of the brain parenchyma.

## 2 Materials and methods

### 2.1 Literature review

Numerous configurations have been suggested in previously published studies to measure the mechanical properties of soft biological tissues. We will focus here on the devices and publications referring to in-vivo measurements. Indeed, it has clearly been shown [Kerdok et al., 2006][Ottensmeyer, 2002][Gefen & Margulies, 2004] that the mechanical behavior of soft tissue can differ significantly between in-vivo and ex-vivo conditions, for a number of reasons, including the vascularization of the tissue. It is therefore of primary importance to provide in-vivo patient-specific measurements in order to accurately simulate brain-shift in Computer Assisted Medical Intervention.

The various excitation methods range from suction to ultrasound with a variety of devices in each category. Aspiration is probably the most widely used technique. Starting from the pioneering work by Grahame [Grahame & Holt, 1969], several authors proposed suction cups differing mostly in the way the aspirated height is measured (optically [Kauer et al., 2002], using ultrasound [Diridollou et al., 2000]) or by their ability to accurately measure the dynamic response. Other excitation methods include indentation [Ottensmeyer, 2001][Carter et al. , 2001][Samur et al. , 2007] using a handheld or robotic indenter, and torsion [Agache et al. , 1980] or ballistometer [Jemec et al. , 2001] which consists of striking the tissue with a known mass and a known force. Ultrasound measurement [Chen et al. , 1996][Gennisson et al. , 2004] is another method linked with the emerging field of elastography. An alternative way of compressing the tissue was proposed by Brown et al. under the form of an endoscopic grasper [Brown et al. , 2003].

Several devices have even become commercial products designed specifically for the dermatology market. Suction apparatus [Cutometer from Courage & Khazaka, Germany, or Dermaflex from Cortex Technology, Denmark] are widely used to measure skin elasticity and are now routinely used for pathology diagnosis. In the same field of application, torque measurement and ballistometer are also available commercially [Diastron Ltd, UK]. In terms of in-vivo measurements, there are two categories available: the instruments usable on external tissues, and those usable during surgery. In the first category, as already mentioned, numerous applications and publications refer to the metrology of skin elastic properties. The most widespread method is aspiration because it is non invasive and proved to be quite reliable. The rate controlled indenter of Pathak et al. [Pathak et al. , 1998] used on residual limb tissue is an example of the first category as is the minimally invasive instrument of Ottensmeyer [Ottensmeyer, 2001]. The second category of per-operative measurement device however, has much fewer candidates. If we restrict our overview to human organs, to our best knowledge, only the indenter of Carter et al. [Carter et al. , 2001] has been used on the human liver and the aspiration device of Vuskovic [Vuskovic, 2001][Kauer et al. , 2002] was tested on the uterine cervix during surgery and more recently on the liver [Nava et al., 2007].

Most of the papers rarely address the important issue of sterilization to any great depth of detail. Sterilization is carried out using very rigorous processes such as a) steam under pressure, b) heat or c) chemicals under a liquid, gaseous or plasma form. The fragile parts of a measurement device can be easily damaged under these conditions, especially electronic parts such as sensors, actuators or circuitry which are not very resistant to such severe environments. Moreover, it is not only the parts in close contact with the operating field that have to be fully sterilized, but also every piece of the instrument which could come into contact with any projection of liquid during surgery. The sterilization process is thus a major concern when considering any measurement apparatus in a real operation. Carter et al. used a sterile sleeve in addition to a removable and sterilizable tip that is effectively in contact with the patient. A recent paper by Samur et al. [Samur et al. , 2007] mentions the sterilization as a key issue. This problem was taken into account at the starting point of the design process of their robotic indenter. They have not yet used this device for human tissue experiments. In the aspiration device on Nava et al. [Nava et al.,2007], the electronic parts are not sterilized but are integrated in the system and the authors mention that they are never in direct contact with the patient.

Moreover, compared to other operations, brain surgery is by far the most demanding in terms of sterilization conditions. In most of the hospitals, prion specifications require the most severe sterilization conditions that sometimes prohibit any other method than pressurized steam. For example sterilization through a temperature of 134° C for 20 minutes completely prohibits the use of any electronic device in contact or in close vicinity with the operating field. This major constraint drove us to the choice of a very simple measurement method with no electronic part involved. The device measuring head that will be described in the following section can almost be viewed as an ancillary tool which could eventually be disposable.

### 2.2 Description of the device

The simplicity and robustness of the aspiration method is very attractive. In order to avoid any electronic component in the measuring head, we designed a transparent suction cup, or pipette, coupled to a 45° mirror that allows us to photograph the deformed tissue using a standard digital camera. A schematic drawing is given in Figure 1.

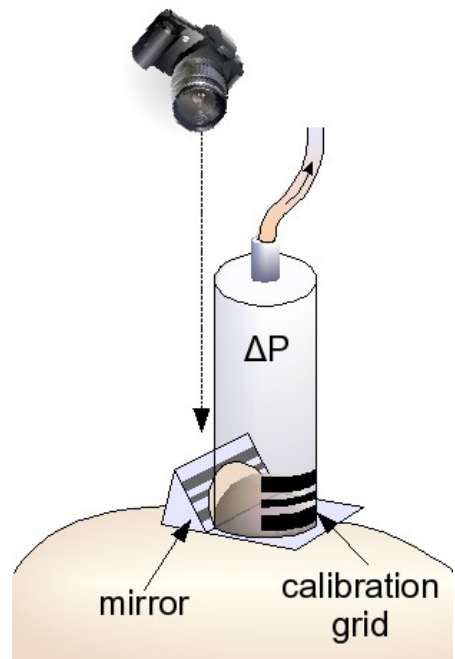

Figure 1: Schematic of the aspiration device.

The top of the cup has a Luer lock connector. A single use Luer lock tube is plugged into the pipette connector. On its other side, the tube is connected to a manometer and a vacuum pump via a T shape Luer-lock connector equipped with taps. In our experiment, a simple syringe acts as a vacuum generator. Negative pressure control is achieved manually, step by step. This simple set-up does not allow for dynamic pressure control and dynamic tissue response measurements. All measurements are therefore considered to be static. Depending on the stiffness of the tissue to be characterized, the maximum negative pressure applied can vary from around 25 mbar for the brain tissue to around 120 mbar for the skin.

A visible calibration grid is placed opposite to the mirror, in the field of view of the camera. In the resulting image, it is viewed in the background of the deformed tissue. It therefore allows an easy measurement of the aspirated height. Note that the practical construction of our current setup leads to an intrinsic "dead zone" in the measurement of small deformations below 1.5 mm that prevents us to get more complete information in the small negative pressure range for very soft materials like brain tissue. A typical picture acquired during measurement is shown in Figure 2. The irregular spacing of the calibration grid provides a very effective method of improving the location of the deformed tissue.

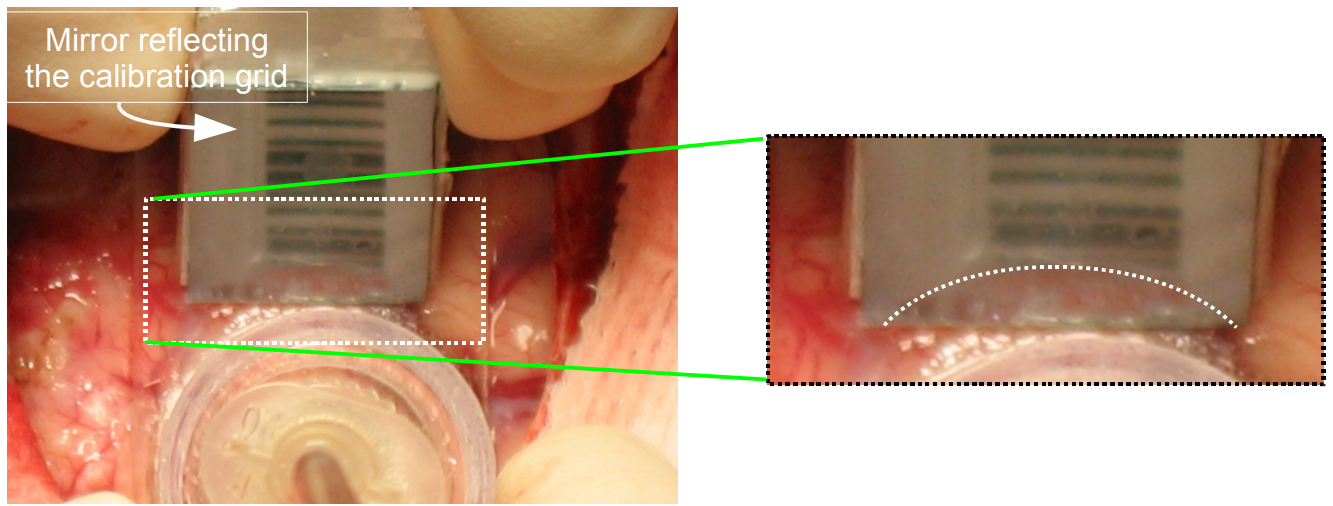

*Figure 2: Close-up of the aspiration device viewed from the top (camera point of view). Note the pattern of the reflected calibration grid.*

The current version of the suction cup is made of a transparent polyethylene plastic. The mirror is made of a rectangular piece of a polished silicon wafer. Although less reflective, it showed much better resistance to the sterilization process than an aluminum counterpart. In this feasibility study we decided to just use ethylene oxide gas sterilization at low temperature. This type of sterilization was mandatory for our plastic cup that was unable to resist to the high temperature steam process. In the future model of the device, only glass and metal will be used; they are the only materials able to withstand the 134° C steam atmosphere for 20 minutes.

**2.3 Experimental protocol**

An adult patient with a brain tumor (*Glioma*), located in the subcortical frontal lobe area, consented to participate to this in vivo study in accordance with the ethical procedure of the INCA (Instituto de NeuroCirugia Asenjo).

The brain tissue deformation measurements took place in the frontal lobe, on the superficial part of the brain tissue (Figure 4) in a non-eloquent area. The clinical procedure is not in any way affected or obstructed by the measurements. No adverse effects were observed due to this measurement nor did we see any change in the brain tissue after the measurement.

Once opened the *Dura Mater*, the suction cup and the Luer-lock tube and locks are taken out from their sterile bag. They then remain in the sterile area at all time. The surgeon connects one extremity of the Luer-lock tube to the suction cup and drops the other out of the sterile area. A first surgeon's assistant then connects the tube to the manometer which has been already plugged in to the vacuum pump. Once the experimental setup is ready, the surgeon presents the suction cup in its final position without touching the brain. During this phase, a second surgeon's assistant prepares the orientation and zoom of the digital camera. This part is important and difficult as the orientation of the camera has to be as parallel as possible to the orientation axis of the cup. For the initial tests, we used a standard 5Mpixels digital camera (Canon PowerShot A95, Canon, Japan). Note that this camera model has a variable angle screen which allows an easier adjustment of the orientation. Once the orientation and zoom of the camera is set, the surgeon puts the cup in contact with the brain tissue. The suction measurements can then start under his supervision. When a visible deformation occurs, the negative pressure is kept constant and the second assistant shoots a picture while the first assistant notes the negative pressure value. The procedure is repeated to obtain other static measurements for larger depression values. The suction cup is then disconnected and sent back to sterilization while the Luer-lock tube is disposed of. The camera, manometer and the vacuum pump never enters the sterile environment and can be directly taken out of the surgical room.

In our experiment, the aspirated height is kept to a safe range of a few millimeters and no lasting deformations of the brain were observed a few seconds after the measurements were stopped. The whole procedure took between 10 and 15 minutes which is not a major drawback as brain surgery can often last for over 4 hours.

**2.4 Finite element modeling and resolution of the inverse problem**

The aspiration experiment itself does not provide the constitutive law of the material. Indeed, the measurements only give the relationship between the local depression applied to the external surface of the body and the resulting displacement. To get the constitutive law from this aspiration experiment, i.e. the global relationship that can be assumed between strain and stress inside the body, an optimization algorithm based on an "analysis by synthesis" strategy was elaborated. It consisted in a four step loop: (1) assume a given constitutive law with a given set of parameters, (2) build and simulate a Finite Element Analysis (FEA) of the aspiration experiment, (3) compare the simulations provided by this FEA with the aspiration measurements in the least square sense, (4) from this comparison deduce better values of the constitutive law parameters in order to improve the FEA simulation/measurement fit. This loop was continued until the comparison carried out in (3) gave satisfactory results. Minimization was computed using an advanced zero-order method, which only required the dependent variable values, and not their derivatives (Subproblem approximation method, Ansys). Being an ill-posed inverse problem, no uniqueness of the solution is guaranteed. However, a careful check has been performed in order to avoid local minima during the optimization process. For example, the reliability of the optimized parameters was checked by changing the initial parameter values and testing the stability of the solution. The FEA were performed using Ansys software (Ansys 10 software, Ansys, Inc., Cannonsburg, PA).

Following previous finite element modeling of the suction experiment [Boudou et al. , 2006], the brain was modeled by a thick circular slice of radius $a$ and thickness $h$ while the pipette was described by a rigid hollow cylinder of internal and external radii $Ri$ and $Re$ respectively (Figure 3).

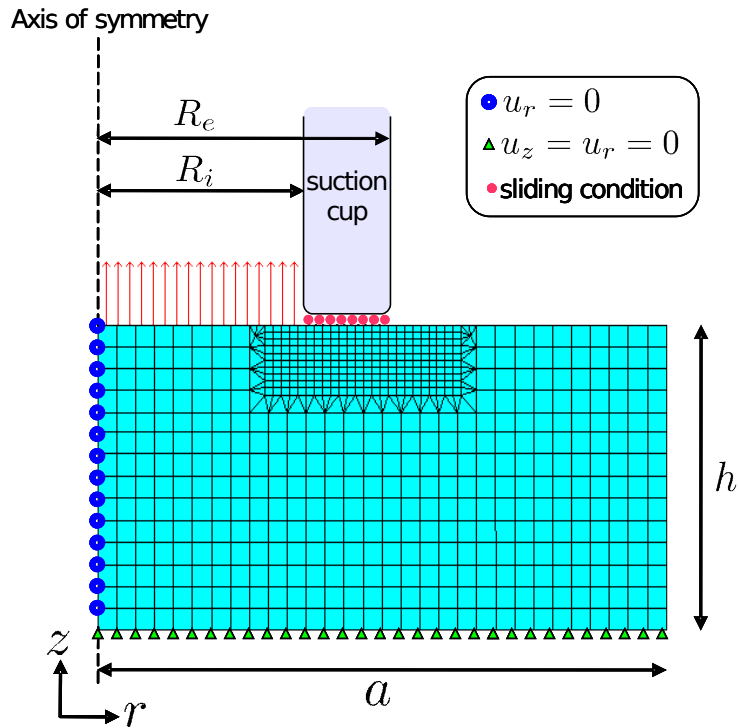

*Figure 3 : Finite element model and boundary conditions of the aspiration experiment*

*$R_i$=5mm; a=50mm, h=50mm*

In order to avoid edge effects, the pipette radius ($R_i$=5mm) was kept small compared to the sample section: in all simulations, the ratios $a/R_i$ and $h/R_i$ have thus been fixed to a constant value equal to 10. Taking advantage of the axi-symmetric geometry of our problem, we reduced its mechanical study to a two-dimensional structural analysis.

We meshed the sample with approximately 2,000 elements, each element being defined by eight nodes. The mesh has been refined in the neighborhood of the aspirated region to get a proper accuracy of the computed solution. The interfacial area between the pipette and the sample was specifically meshed with contact elements in order to insure that the aspirated brain tissue slides without friction over the tip of the pipette. The frictionless condition is very close to what was observed during our experiments where brain surface proved to be very slippery. The following conditions were imposed on the sample boundaries: (i) a controlled suction pressure ΔP was applied on the sample section inside the pipette, (ii) zero normal displacement conditions were applied along the sample section belonging to the axis of symmetry and (iii) zero displacement was imposed on the basal surface (see Figure 3). Free boundary conditions were assumed for all other sample surfaces.

## 3 Results

### 3.1 Measurements

The brain tissue deformation measurements took place in the frontal lobe. No initial compression was needed in our experiments. Six negative pressures were applied from 0 mbar to 25 mbar. An image was shot for each corresponding deformation. On each image, the maximum height of the aspiration was manually measured in pixels relatively to the calibration grid, thus providing a height in mm. Once the pressure was reset, the brain tissue went back to their original configuration.

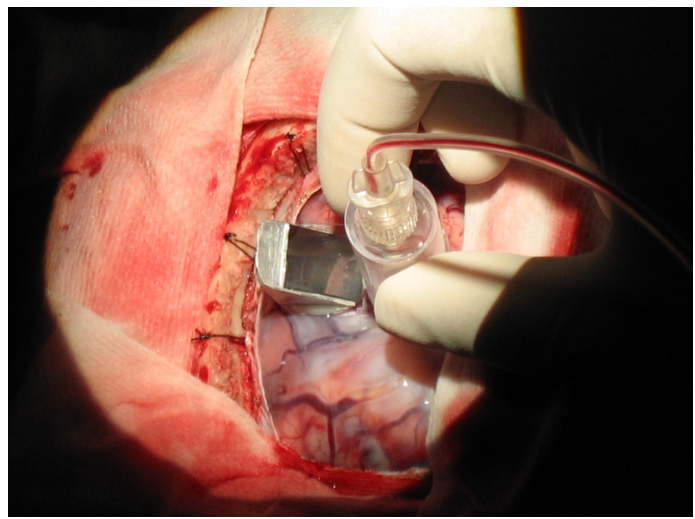

*Figure 4: View of the brain pipette aspiration experiment in intra-operative conditions*

**3.2 Estimation of a constitutive law**

In light of the nonlinear stress–strain relationships given by the pipette aspiration measurements, the brain medium was assumed to be an isotropic homogeneous and hyperelastic continuum. Among the various strain-energy functions that can describe this pattern of mechanical response we selected a two-parameter Mooney-Rivlin strain-energy function W given by the following analytical expression:

$$W = a_{10}(I_1 - 3) + a_{30}(I_1 - 3)^3 \qquad (Eq.\ 1)$$

where $a_{10}$ and $a_{30}$ are the two material constants (in Pa) and $I_1$ is the first invariant of the right Cauchy–Green strain tensor C with $I_1 = \text{Trace}(C)$. We observed that this formulation of the Mooney-Rivlin law using coefficient $a_{30}$ instead of the more classical formulation with coefficient $a_{20}$ (i.e. with $W = a_{10}(I_1 - 3) + a_{20}(I_1 - 3)^2$) provided a better fit of the data. Different values of the Poisson ratio can be found in the literature ranging from 0.4 [Skrinjar et al., 2002] to 0.5 [Taylor & Miller, 2004]. However, like most recent studies [Clatz et al., 2005][Ferrant et al., 2002] [Takizawa et al., 1994], we used a nearly incompressible continuum material with a Poisson's ratio ν = 0.45 to model the brain medium.

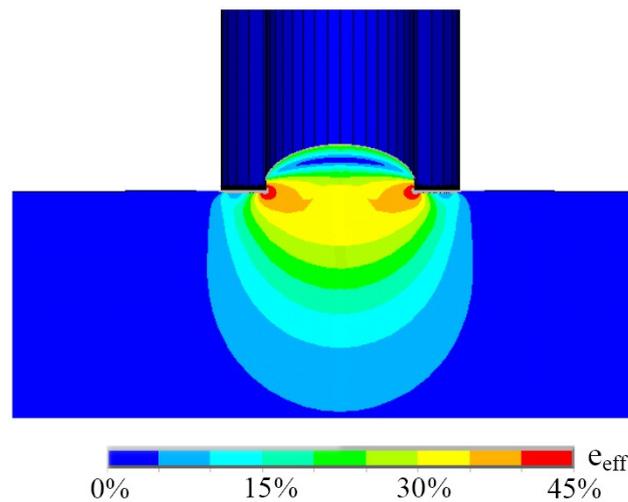

*Figure 5: Finite element calculation of the deformation under a 25 mbar negative pressure using the*

*experimentally determined non linear constitutive law. The corresponding Von Mises strain values are provided.*

Figure 5 plots one FEA simulation of the aspiration experiment with the maximal negative pressure value used with the experimental measurements. As can be observed, a maximal strain level of 45% is simulated which justifies the non-linear *large deformation* framework assumed with the Mooney-Rivlin constitutive law. A good fit of the data (Figure 6) was obtained for $a_{10}$ = 0.24 kPa and $a_{30}$ = 3.42 kPa, indicating that the mechanical non-linear response of the brain was well described over the entire strain range by this pair of values.

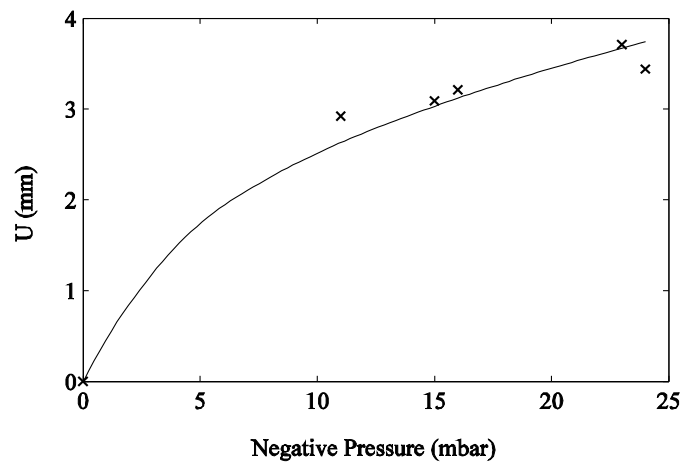

*Figure 6: Best fit of the experimental data (measured maximum height of the aspiration U vs. the applied negative pressure P) and simulated data (non-linear Mooney-Rivlin constitutive law with a10=240 Pa, a30=3420 Pa and ν=0.45).*

## 4 Discussion

### 4.1 Choice of the aspiration device

For a large number of elasticity measurement devices, the positioning of the measuring head on the tissue is very critical. Most of the measuring devices require a precise positioning with respect to the tissue, e.g. the indenter in [Ottensmeyer, 2001]. This is often very difficult to assure in a real operating situation, even when the patient is anesthetized, because numerous movements can occur (for example respiratory movements) that superimpose on the movement of the indenter tip. The relevant signal can be hard to extract even after post-processing. This underlines the lack of stability of the reference landmark. The same holds partly true for relatively heavy hand-held tools [Nava et al. , 2003] [Carter et al. , 2001]. One way to address this issue is to use a robotic instrument [Samur et al. , 2007]. Our approach is different since our measurement head is very light (20g); once in contact with the tissue, the negative pressure generates a suction force that sticks the tissue to the pipette, which does not need to be held anymore. This means that the measurement head is referenced with respect to the tissue itself and moves with it. Only the relative displacements are measured without influence of the weight or of the force applied by the operator. However, the drawback is that it can only be used for "open" surgery. It is therefore not applicable to a laparoscopic approach.

### 4.2 Constitutive law

The main aim of this paper is not to describe a universal constitutive law for the brain elasticity. We mostly want to disclose here the results of a difficult experiment and provide for the first time *in vivo* data on the human brain elasticity. We would like to emphasize that the measurement tool and the way the inverse problem is dealt with are somewhat disconnected. The same holds for the degree of sophistication of the constitutive law. The same set of data could be used to fit a small deformation law or a more sophisticated elastic model.

### 4.3 Main sources of measurements errors

The inversion process being based on the optimisation of a highly non linear function, the inference of the error on the estimated constitutive law parameters cannot be performed in a straightforward way. No exact mathematical formulation exists in this case. Nevertheless, we tried to estimate what are the major contributors to measurement errors.

The measured tissue deformations are generally in the 1-3 mm range. The precision of the extraction from the image is estimated to 3 pixels which corresponds to 0.01 mm, i.e. less than 1% of the deformation. Another main source of the measurement error is the orientation of the camera during the image capture (the optical axis of the camera has to be aligned with the pipette cylinder axis). From the acquired pictures, we could estimate that this angular error was maintained by the operator below 3°. Using simple geometrical optics considerations, we computed that the error induced by the misalignment on the measurement of the aspirated tissue height is 1% per degree. It means that the uncertainty caused by the sole misalignment is estimated to 3%. However, although we can evaluate this particular source of error quantitatively, it is in fact a minor contributor to the overall error budget. We found that in our experimental conditions, the major cause of error was due to the accuracy of the applied negative pressure and in particular to its poor synchronization with the image capture since the pressure reading and image capture were done by two different operators.

### 4.4 Perspectives

We identified some problems and their respective solutions will be implemented in the second phase of this study. The first one concerns sterilization. As plastic was used, autoclave-proof procedure can not be carried out, unless the device is considered as a disposable ancillary. Considering a reusable device, we plan to develop a glass cup which will support autoclave sterilization, it will certainly be slightly heavier than a plastic device, but we think that it can be kept very compact and consequently light. Water vapour was sometimes observed on the surface of the cup; a special hydrophilic treatment of the glass should solve this issue. The accuracy of the applied negative pressure could also easily be improved by using a more stable vacuum generator and an automated synchronization of the camera acquisition with the pressure application.

**Acknowledgement**

The authors would like to acknowledge Marc Poulet and Catherine Guimier from the sterilization group at Grenoble University Hospital for their help in the understanding of the sterilization process. The nurses and technical staff of the operating room and sterilization unit of the Instituto de Neurocirugía Asenjo (INCA) are also gratefully acknowledged. The authors are finally indebted to Francisco Galdames for his assistance and to Professor Jacques Ohayon for discussions about suction experiments and model inversion.